\documentclass[aps,prd,reprint,superscriptaddress,nofootinbib]{revtex4-1}

\usepackage{amsmath}
\usepackage[dvipsnames]{xcolor}
\usepackage{graphicx}
\usepackage{hyperref}
\usepackage{braket}
\hypersetup{
 bookmarksnumbered=true,
 colorlinks=true,
 linkcolor=[rgb]{0.098,0.098,0.439},
 citecolor=[rgb]{0.098,0.098,0.439},
 urlcolor=[rgb]{0.098,0.098,0.439}
  }

\begin{document}

\title{Subfrequency search at a laser as a signal of dark matter sector particles}
\author{Hye-Sung Lee}
\affiliation{Department of Physics, KAIST, Daejeon 34141, Korea}
\author{Jiheon Lee}
\affiliation{Department of Physics, KAIST, Daejeon 34141, Korea}
\author{Jaeok Yi}
\affiliation{Department of Physics, KAIST, Daejeon 34141, Korea}
\date{January 2022}

\begin{abstract}
We introduce a new method to search for the dark matter sector particles using laser light.
Some dark matter particles may have a small mixing or interaction with a photon. High-power
lasers provide substantial test grounds for these hypothetical light particles of exploding interests
in particle physics. We show that any light source can also emit a subfrequency light as a new
physics signal. Searching for this subfrequency light at a laser can be a simple and effective way to
investigate the new light particles, even in tabletop optics experiments.
\end{abstract}

\maketitle

%%%%%%%%%%%%%%%%%%%%%%%%%%%%%%%%%
\section{Introduction} 
%%%%%%%%%%%%%%%%%%%%%%%%%%%%%%%%%
One of the characteristic trends in particle physics research in the past decade or two is investigating the very light dark matter sector particles.
While the standard model (SM) is the utmost successful model of the unveiled particles and their interactions, cosmological observations indicate that 95\% of the Universe consists of nonluminous matter and energy, which cannot be explained by the SM \cite{Planck:2018vyg}.
The mysterious dark sector may have various kinds of particles as the SM sector, and the two sectors may have a feeble interaction channel referred to portals \cite{Essig:2013lka}.

This tiny size of interactions allows extremely light mass for the dark matter sector particles that often require an ingenious strategy to probe.
It brings the other fields such as optics and condensed matter physics into a new physics search originating from the dark matter sector, permitting genuinely interdisciplinary studies \cite{ARIADNE:2017tdd,ADMX:2021nhd,CAST:2020rlf,Trickle:2019ovy,Ehret:2010mh,Spector:2016vwo,Robilliard:2007bq,GammeVT-969:2007pci,Afanasev:2008fv,OSQAR:2015qdv,Cameron:1993mr}.

Among the popular dark sector particles are an axion ($a$) and a dark photon ($\gamma'$).
They are exceptional in the sense that they mix or couple to a photon ($\gamma$), which chiefly distinguishes the dark sector from the visible sector made of the SM particles.
These particles have been searched for using this connection to the light in the lab experiments as well as in the astrophysical observations.
(See Refs.~\cite{ParticleDataGroup:2020ssz,Kim:2008hd,physofdp} for the reviews on the related subjects.)

\begin{figure}[b]
\centering
\includegraphics[width=0.4\textwidth]{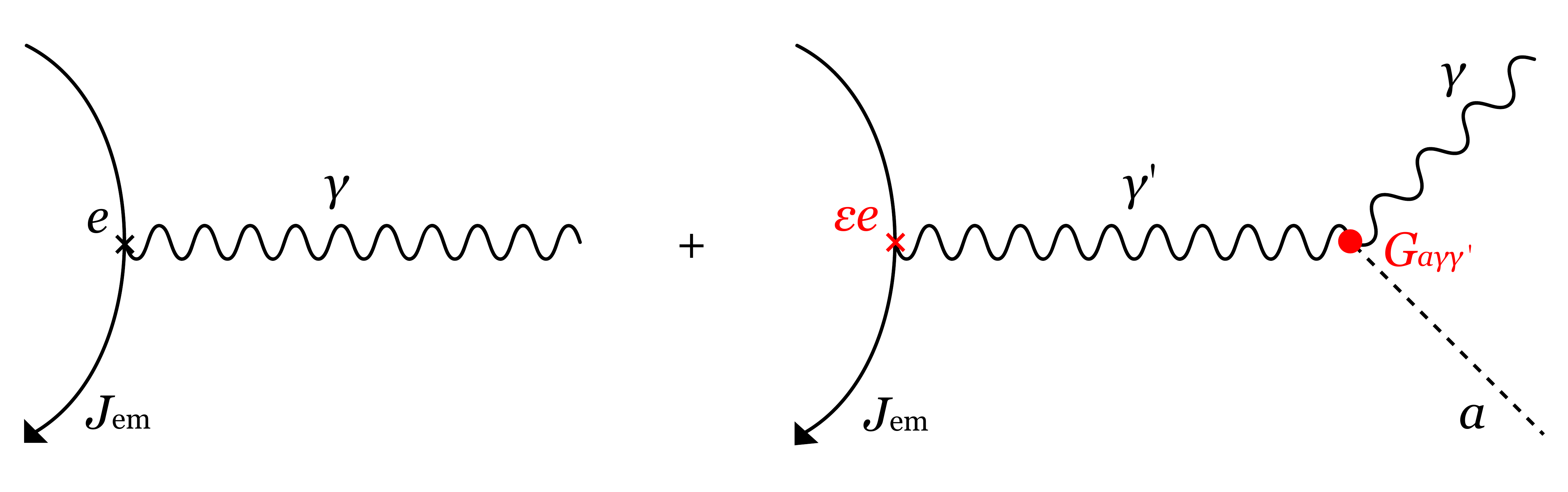}
\caption{Diagrams showing how the subfrequency photon can be emitted through the combined dark sector portals ($\varepsilon$ and $G_{a\gamma\gamma'}$) when there is a source of a photon.}
\label{fig:diagrams}
\end{figure}

In this paper, we make one key observation regarding a photon and the dark matter sector and study its implication in one experimental setup, which is a laser experiment.
Our observation is following.
Whenever there is a source that emits light, the same source may also emit subfrequency light due to the connections of a photon to some dark sector particles.
A dark photon may couple to the electromagnetic current because of the kinetic mixing with a photon.
It decays to an axion and a lower energy (subfrequency) photon, which can be the new physics signal.
This is a very general property, which is depicted in Fig.~\ref{fig:diagrams}.
As there is a great number of sources of light, the potential applications of this possibility are far reaching.
We find the optical laser system is a simple and effective way to study this new physics.
Although there have been a few laser experiments to search for new physics, this one is qualitatively different from any of them in generating and detecting the new physics signal.

The rest of this paper is organized as followings.
In Sec.~\ref{sec:newphysics}, we describe the relevant Lagrangian for the new physics scenario.
In Sec.~\ref{sec:subfrequencylight}, we describe how the subfrequency light can be produced from the dark photon decay.
In Sec.~\ref{sec:kinematics}, we describe the kinematics of the relevant process.
In Sec.~\ref{sec:newlaserexperiment}, we propose a new laser experiment and the sensitivity of the experiment.
In Sec.~\ref{sec:discuss}, we discuss the general properties of the new idea and experiment.
In Sec.~\ref{sec:summary}, we summarize our findings and depict outlooks.

%%%%%%%%%%%%%%%%%%%%%%%%%%%%%%%%%
\section{New physics}
\label{sec:newphysics}
%%%%%%%%%%%%%%%%%%%%%%%%%%%%%%%%%

The SM particles do not carry a charge of the dark sector gauge symmetry $U(1)_\text{dark}$.
The vector portal ($\varepsilon$) \cite{Holdom:1985ag} is a dimensionless parameter, which is a small kinetic mixing between the $U(1)_\text{QED}$ and the $U(1)_\text{dark}$.
Through this $\gamma$-$\gamma'$ mixing, the dark photon can couple to the electromagnetic current with a suppressed coupling constant $\varepsilon e$ in contrast to the electromagnetic coupling constant $e$.

The dark axion portal ($G_{a\gamma\gamma'}$) \cite{Kaneta:2016wvf} is a nonrenormalizable $a$-$\gamma$-$\gamma'$ vertex, whose actual size depends on the details of the underlying theory.
(By the axion, we mean the axionlike particle in general, not necessarily related to the strong $CP$ problem \cite{Peccei:1977hh,Weinberg:1977ma,Wilczek:1977pj}.)
Many implications of this portal have been found in various contexts \cite{Choi:2016kke,Kitajima:2017peg,deNiverville:2018hrc,Arias:2020tzl,Deniverville:2020rbv,Hook:2021ous,Ge:2021cjz}.

The relevant part of the Lagrangian in the mass eigenstate is
\begin{equation}
\mathcal{L} \sim - \big(A^\text{phys}_\mu + \varepsilon {A'}^\text{phys}_{\mu}\big) J_\text{em}^\mu + \frac{G_{a\gamma\gamma'}}{2} aF^\text{phys}_{\mu\nu}\tilde{F}_\text{phys}'^{\mu\nu}\label{eq:lagrangian}
\end{equation}
where $J_\text{em}^\mu$ is the electromagnetic current,
$F^\text{phys}_{\mu\nu} = \partial_\mu A^\text{phys}_\nu - \partial_\nu A^\text{phys}_\mu$ is the field strength tensor for the photon field $(A_\mu^\text{phys})$, and ${F'}^\text{phys}_{\mu\nu}$ is for the dark photon field $({A'}_\mu^\text{phys})$.
For the dark photon heavier than the axion, $\gamma' \to a \gamma$ decay can occur with the $G_{a\gamma\gamma'}$.
Although it is possible to consider the axion decay into a pair of photons in the presence of the axion portal ($G_{a\gamma\gamma}$) allowing $\gamma' \to a \gamma \to \gamma \gamma \gamma$, we treat things as model independent as possible and assume the axion portal is zero in this paper. In this sense, the number of photon signals in our analysis will be conservative.
The full Lagrangian and the extended Maxwell's equations in the presence of various dark sector portals can be found in Ref.~\cite{Huang:2018mkk}.

%%%%%%%%%%%%%%%%%%%%%%%%%%%%%%%%%
\section{Subfrequency light from the dark photon decay}
\label{sec:subfrequencylight}
%%%%%%%%%%%%%%%%%%%%%%%%%%%%%%%%%
In the following, we use the number flux of laser photon $N_{\gamma}$, defined as $P/\omega$, where $P$ is the laser power and $\omega$ is the frequency.
At the laser, a dark photon with mass $m_{\gamma'}$ can be produced through the suppressed coupling to the electromagnetic current.
Alternatively, this dark photon can be understood in terms of the photon-dark photon oscillations in a similar fashion to the neutrino flavor oscillations, and it has been searched for in the light shining through the wall (LSW) experiments \cite{Ehret:2010mh,Spector:2016vwo,Robilliard:2007bq,GammeVT-969:2007pci,Afanasev:2008fv,OSQAR:2015qdv,Cameron:1993mr}. The largest mass of the dark photon that we can produce from the optical laser is the laser frequency of the eV scale ($m_{\gamma'} < \omega$).

In the presence of a lighter axion, the dark photon may decay into a photon and an axion with a decay rate
\begin{equation}
\Gamma \equiv \Gamma_{\gamma^{\prime} \rightarrow a \gamma} = \frac{G_{a\gamma\gamma^{\prime}}^2}{96\pi}m_{\gamma^{\prime}}^3\bigg(1-\frac{m_a^2}{m_{\gamma^{\prime}}^2}\bigg)^3 .
\label{eq:decayrate}
\end{equation}
The photon produced from this dark photon decay, which we call the signal photon, has a smaller frequency than the original laser frequency. 

In order to demonstrate the production of the signal photon, we investigate the photon-dark photon oscillation together with the decay of massive dark photons. We define the massless eigenstate as a photon mass eigenstate $\ket{\gamma}$ and massive eigenstate as a dark photon mass eigenstate $\ket{\gamma^{\prime}}$. The mass eigenstates propagate as plane waves of the form 
\begin{align}
    \ket{\gamma(t,L)}& = e^{-i(\omega  t - \omega L)} \ket{\gamma} \, , \\
    \ket{\gamma^{\prime}(t,L)} &= e^{-i(\omega t - p L)} e^{-m_{\gamma^{\prime}}\Gamma t /2 \omega} \ket{\gamma^{\prime}} \, ,
\end{align}
where $p = \sqrt{\omega^2-m_{\gamma^{\prime}}^2}$ is the momentum of the dark photon.
The Lorentz factor $\gamma = \omega / m_{\gamma'}$ reflects the time dilation of the decay. Here, $t$ and $L$ are measured in the lab frame.

When the photon is produced in the laser, it is produced in a flavor eigenstate $\ket{A}$ and its orthogonal pair is $\ket{X}$. They are related to the mass eigenstates $\ket{\gamma}$ and $\ket{\gamma^{\prime}}$ by $2\times2$ unitary matrix.
\begin{equation}
    \begin{bmatrix}
     A \\ X
    \end{bmatrix} = \begin{bmatrix}
    \sqrt{1-\varepsilon^2} & \varepsilon \\ -\varepsilon & \sqrt{1-\varepsilon^2}
    \end{bmatrix} \begin{bmatrix}
    \gamma \\ \gamma^{\prime}
    \end{bmatrix}
\end{equation}
Let our state emitted from the laser be $\ket{\psi(t,L)}$. Since the initial state $\ket{\psi(0,0)}=\ket{A}$, the state after time $t$ and length $L$ of propagation is
\begin{align}
    \ket{\psi(t,L)} & = \sqrt{1-\varepsilon^2}  e^{-i(\omega  t - \omega L)} \ket{\gamma} \nonumber \\
    & +  \varepsilon e^{-i(\omega t - p L)- m_{\gamma^{\prime}} \Gamma t / 2\omega} \ket{\gamma'} \, .
\end{align}
We can obtain a probability that the initial $\ket{A}$ turns into $\ket{X}$ after propagation as
\begin{align}
    \mathcal{P}_{A \rightarrow X } (L) & =  |\braket{X|\psi(t,L)}|^2 \nonumber \\
    & = \varepsilon^2  \bigg[1+ e^{-\frac{m_{\gamma'}\Gamma t}{\omega} } -2  e^{-\frac{ m_{\gamma^{\prime}}\Gamma t}{2 \omega}} \cos (L \Delta p ) \bigg]  \nonumber \\
    & = \varepsilon^2  \bigg[1+ e^{-\frac{m_{\gamma'}\Gamma L}{p} } -2  e^{-\frac{ m_{\gamma^{\prime}}\Gamma L}{2 p}} \cos (L \Delta p ) \bigg] 
\end{align}
where $\Delta p = \omega - p = \omega -\sqrt{\omega^2-m_{\gamma^{\prime}}^2}$ and the terms with $\mathcal{O}(\varepsilon^4)$ are neglected. (We assume small kinetic mixing.) We used $L = v t$ and $p = \omega v$ in the last line.
Likewise, a simple calculation gives the probability that $X$ results in decay products.
\begin{align}
    \mathcal{P}_{X\rightarrow \text{decay}} (L) & = 1- \mathcal{P}_{X\rightarrow A }(L)- \mathcal{P}_{X\rightarrow X }(L) \nonumber \\ &  = (1-\varepsilon^2)\left[1 - e^{-\frac{m_{\gamma^{\prime}}\Gamma L}{p}} \right] \label{eq:decay}
\end{align}
Here, the oscillation behavior is suppressed with $\mathcal{O}(\varepsilon^4)$.

As we will describe in Sec.~\ref{sec:newlaserexperiment}, although it is not the only possible option, our proposed setup is basically the LSW-type experiment in which there are two baselines of length $L_1$ and $L_2$ that are separated by a ``wall'' in between. So the laser light initially in the $\ket{A}$ travels the $L_1$, then only the $\ket{X}$ component of the light travels the $L_2$ (as a photon cannot go through the wall) decaying into an axion and a subfrequency photon, which we measure at the end of $L_2$.

Therefore, the flux of the subfrequency signal photon is given as 
\begin{align}
N_\text{sub} &= \mathcal{P}_{A \rightarrow X} (L_1) \mathcal{P}_{X\rightarrow \text{decay}} (L_2) N_\gamma \nonumber \\ & = \varepsilon^2 N_\gamma\nonumber \bigg[1 - e^{-\frac{m_{\gamma^{\prime}}\Gamma L_2}{p}} \bigg] \\ & \times \bigg[1+ e^{-\frac{m_{\gamma'}\Gamma L_1}{p}} -2  e^{-\frac{ m_{\gamma^{\prime}}\Gamma L_1}{2 p}} \cos (L_1\Delta p ) \bigg] .\label{eq:Nsub}
\end{align}

In this paper, we work on a ``simple setup'' described by the following conditions.
\begin{enumerate}
\item[(i)] $m_a \ll m_{\gamma^{\prime}} < \omega$.
\item[(ii)] $\dfrac{m_{\gamma^{\prime}} \Gamma L_i}{\sqrt{\omega^2 - m_{\gamma^{\prime}}^2}} \ll 1$ ~(for $i=1,\ 2$).
\item[(iii)] $L_1 \Delta p \gg 2\pi $.
\end{enumerate}
These are reasonable assumptions for the optical experiment in the laboratory scale for the following reasons.

Condition (i) reflects that the dark photon should be produced from the laser ($m_{\gamma^{\prime}} < \omega$) and decay into a much lighter axion ($m_a \ll m_{\gamma^{\prime}}$) and a subfrequency photon. 
The latter ensures the subfrequency photon has a sizable number of events and a visible light frequency range that can be detected with a typical photon detector, although we will briefly discuss how the non-negligible axion mass ($m_{a} \sim m_{\gamma^{\prime}}$) can change the result in Sec.~\ref{sec:discuss}.

Condition (ii) is the requirement to put the relevant exponential expressions in linear form of $L$ as in Eq.~\eqref{eq:signalratio}, which is convenient for analysis.
Under this condition, the probability of the dark photon decay during the propagation is much less than 1. Using Eq.~\eqref{eq:decayrate}, this can be rewritten as 
\begin{equation}
G_{a\gamma\gamma^{\prime}}^2 \ll 96\pi \frac{\sqrt{\omega^2 - m_{\gamma^{\prime}}^2}}{m_{\gamma^{\prime}}^4} \frac{1}{L_2} \,  . \label{eq:Limit}
\end{equation}
This can be easily satisfied for the typical lab-size experiment scales if $m_{\gamma^{\prime}} \ll \omega$.
The left-hand side of condition (ii) diverges when $m_{\gamma^{\prime}}\sim \omega$, but it hardly occurs because of the finite line width of a laser. Thus only a tiny fraction of the dark photon can have an enhancement effect, even if the $m_{\gamma^{\prime}}$ happens to coincide the laser frequency.

Condition (iii) allows us to neglect the oscillation in $\mathcal{P}_{\gamma\rightarrow\gamma'}$. If $L_1 \Delta p  $ is much larger than $2\pi$, the cosine function in $\mathcal{P}_{\gamma \rightarrow \gamma'}$ is highly oscillatory. The linewidth of laser in the setup can alter $\Delta p$, and its small change can alter the signal size uncontrollably. To treat this oscillation, we take the average value of the cosine function in Eq.~\eqref{eq:Nsub}. Even in the low mass region, this condition can be satisfied by our interested parameter region for $L_1 > 1 \text{m}$.

In the simple setup described above, the signal photon ratio can be obtained by
\begin{equation}
 \frac{N_{\text{sub}}}{N_\gamma}=\frac{K^2}{48\pi}\frac{m_{\gamma^{\prime}}^4}{\sqrt{\omega^2-m^2_{\gamma^{\prime}}}}L
 \label{eq:signalratio}
\end{equation}
where $L=L_2$ and the effective portal coupling $K$ is given as a product of the two portal couplings
\begin{equation}
K \equiv \varepsilon \, G_{a\gamma\gamma'} .
\end{equation}
The fraction of the subfrequency light is given by two model parameters $K$, $m_{\gamma^{\prime}}$, the laser frequency $\omega$, and a geometry factor $L$.
The factor of 2 difference between Eqs.~\eqref{eq:decayrate} and \eqref{eq:signalratio} comes from the fact that both mass eigenstates generate the $X$ state.

%%%%%%%%%%%%%%%%%%%%%%%%%%%%%%%%%
\section{Kinematics}
\label{sec:kinematics}
%%%%%%%%%%%%%%%%%%%%%%%%%%%%%%%%%
Consider the rest frame of a dark photon, which decays into a subfrequency signal photon and an axion.
In our simple setup where an axion is considered nearly massless, both the photon and axion have the same energy $m_{\gamma'}/2$.
Since there is no preference of direction in the rest frame, the probability density ($dp/d\Omega$) that the signal photon heads to a specific solid angle $d\Omega$ is given by
\begin{equation}
\frac{dp}{d\Omega} = \frac{1}{4\pi} , \quad \int \frac{dp}{d\Omega} d\Omega = 1 .
\end{equation}
By denoting the solid angle with the polar angle $\theta_\text{rest}$ and azimuthal angle $\phi_\text{rest}$, we have
\begin{equation}
\int \frac{dp}{d\Omega} \sin\theta_\text{rest} d\theta_\text{rest} d\phi_\text{rest} = \int \frac{\sin \theta_\text{rest}}2  d\theta_\text{rest} = 1, \end{equation}
where we used azimuthal symmetry in the first equality. Using this relation, we can define the probability density function with respect to $\theta_\text{rest}$:
\begin{equation}
    \frac{dp}{d\theta_\text{rest}} = \frac{\sin\theta_\text{rest}}2
\end{equation}

Let the $z$ axis be the direction a dark photon travels in.
In the lab frame, the dark photon has the energy $\omega=\gamma m_{\gamma'}$, where $\gamma=1/\sqrt{1-\beta^2}$ is the Lorentz factor.
Thus, the lab frame boosts with velocity $\beta=\sqrt{1-(m_{\gamma'}/\omega)^2}$.
Then the energy $E$ and polar angle $\theta_{\text{lab}}$ of a signal photon in the lab frame are
\begin{eqnarray}
E &=& \frac{\omega}2 (1+ \beta \cos\theta_{\text{rest}}) , \label{eq:energydispersion1} \\
\tan \theta_{\text{lab}} &=& \frac{\sin\theta_{\text{rest}}}{\gamma (\cos\theta_{\text{rest}} + \beta)} \label{eq:energydispersion2} .
\end{eqnarray}
 
\begin{figure*}[bt]
    \centering
    \includegraphics[width=.45\textwidth]{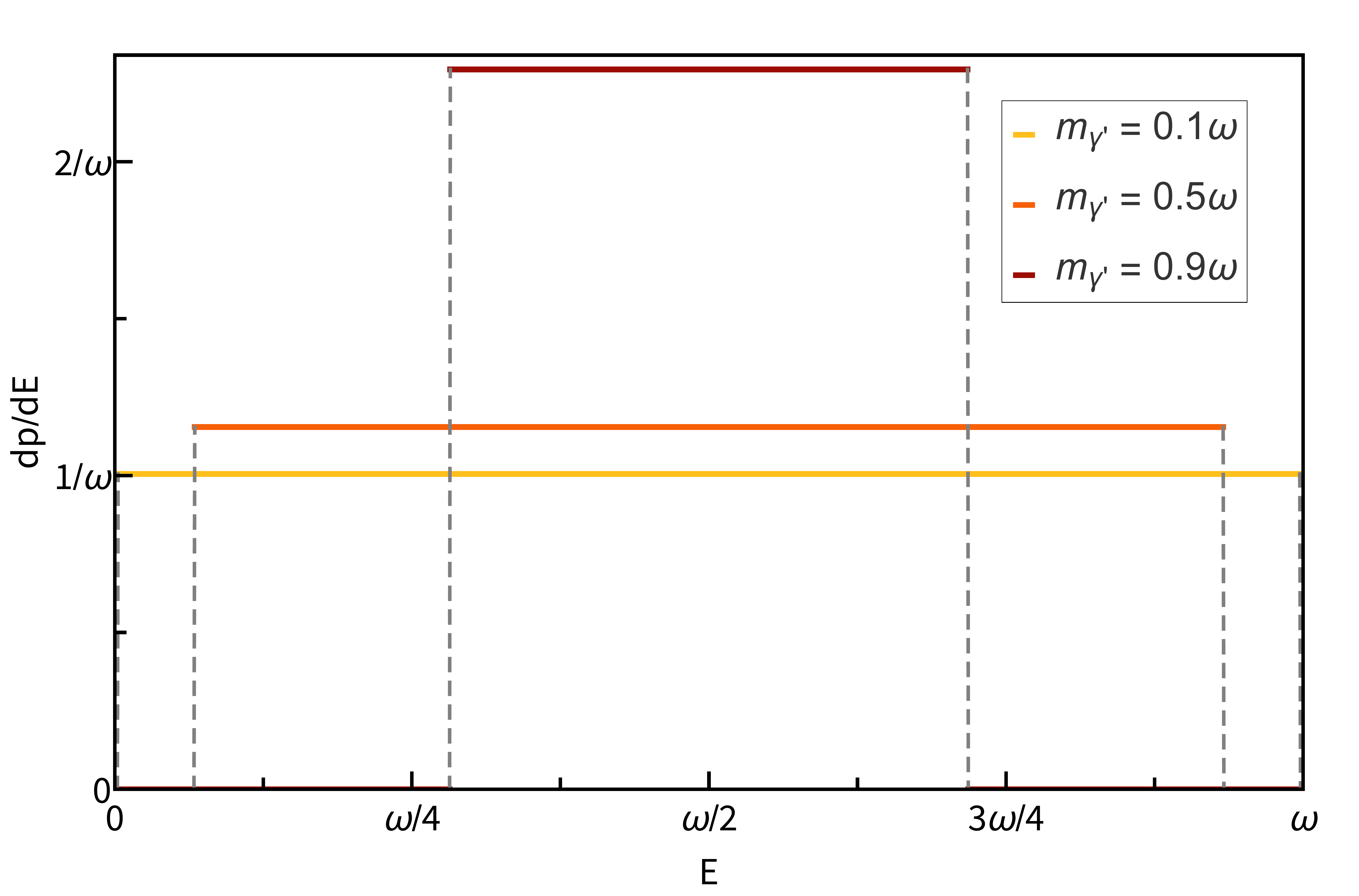} \qquad 
    \includegraphics[width=.45\textwidth]{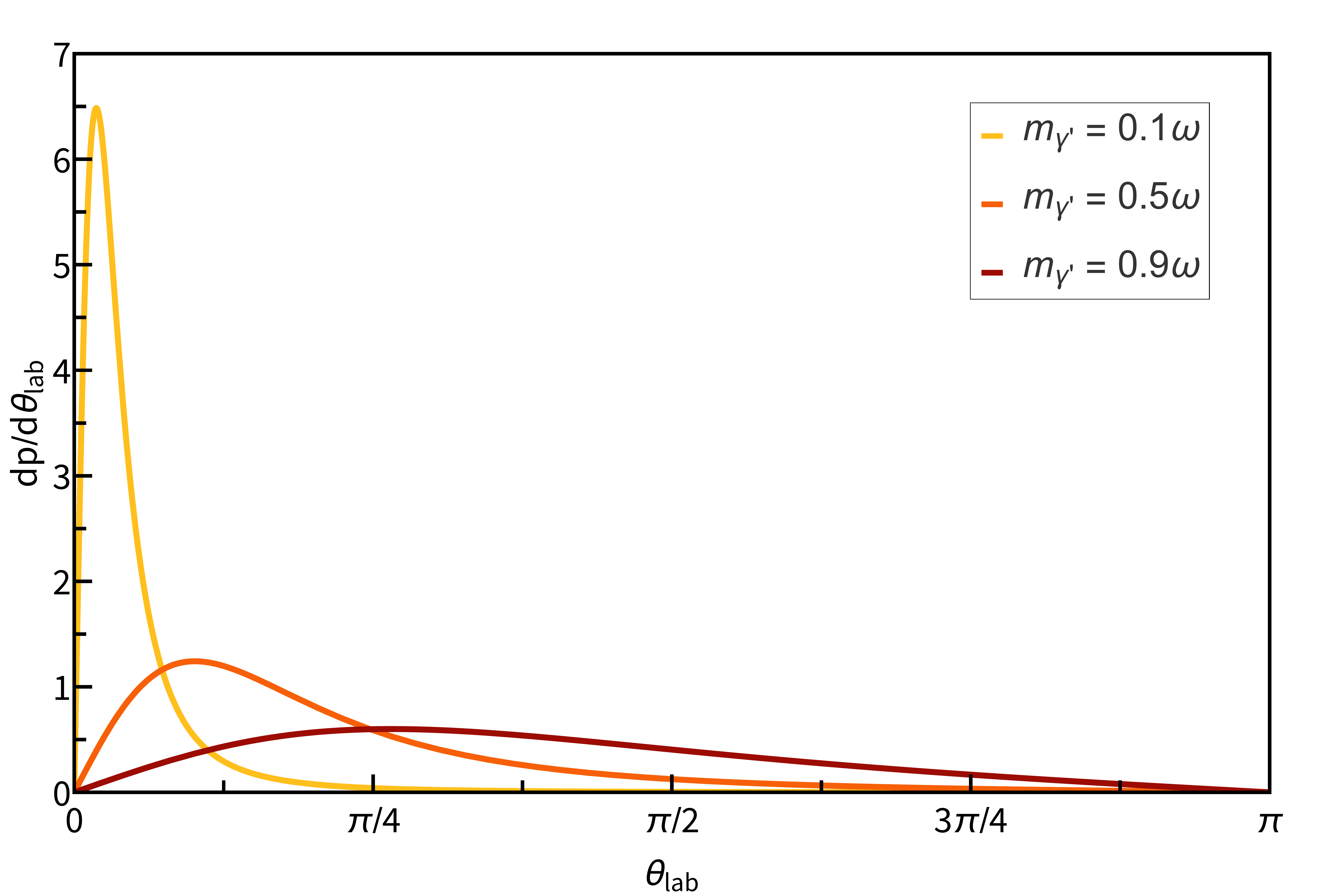}
    \caption{Probability density function with respect to the energy $E$ of the  subfrequency signal photon and the angle in the lab frame $\theta_\text{lab}$. As the dark photon mass $m_{\gamma'}$ increases, the energy of the signal photon is focused around $\omega/2$ but angular distribution gets more disperse.}
    \label{fig:pd}
\end{figure*}

If $\beta \ll 1$ ($m_{\gamma'} \approx \omega)$, then $E$ is not dispersive. But $\theta_{\text{lab}}$ gets close to $\theta_{\text{rest}}$ and the signal photons spread almost isotropically.
On the other hand, $E$ is highly dispersive if $\beta \approx 1$ ($m_{\gamma'} \ll \omega)$.
For a fixed $\omega$, as $m_{\gamma'}$ increases, the dispersion of $E$ decreases, but that of $\theta_{\text{lab}}$ increases. We can check these dispersions by computing the probability density with respect to $E$ and $\theta_\text{rest}$ for each. They are given by
\begin{equation}
    \frac{dp}{dE}  = \frac{dp}{d\theta_\text{rest}} \left| \frac{d\theta_\text{rest}}{dE}\right| , \quad 
    \frac{dp}{d\theta_\text{lab}}  = \frac{dp}{d\theta_\text{rest}} \left| \frac{d\theta_\text{rest}}{d\theta_\text{lab}}\right| .
\end{equation}
These probability distributions and dispersive behaviors are illustrated in Fig.~\ref{fig:pd}.
    
One of the interesting features of the energy dispersion, as we can find in Fig.~\ref{fig:pd}, is that the distribution is uniform. This probability distribution function is given by
\begin{equation}
    \frac{dp}{dE}  = \frac{1}{\sqrt{\omega^2-m_{\gamma'}^2}} ~\text{ for } \frac{\omega}2 (1-\beta) < E <\frac{\omega}2 (1+\beta) .
\end{equation}
The energy dispersion also shows that the frequency of resulting signal is smaller than the laser frequency $\omega$, which justifies our naming ``subfrequency.'' However, one should note that despite the dispersion of $E$, a signal photon whose energy is much smaller than $\omega/2$ is produced very little and in the backward direction as one can see from Eqs.~\eqref{eq:energydispersion1}--\eqref{eq:energydispersion2}.

Because of the angular dispersion of the subfrequency photon, the location and the size of the detector could matter in detecting the signal.
We want to estimate how much signal can be collected in a circular detector of radius $a$ located at $L$ from the dark photon production. Often, a lens is installed in front of a detector to focus the beam into the small detector, then the ``$a$'' can be replaced by the radius of the lens as an effective detector size in the conservative sense.

\begin{figure}[b]
    \centering
    \includegraphics[width=.45\textwidth]{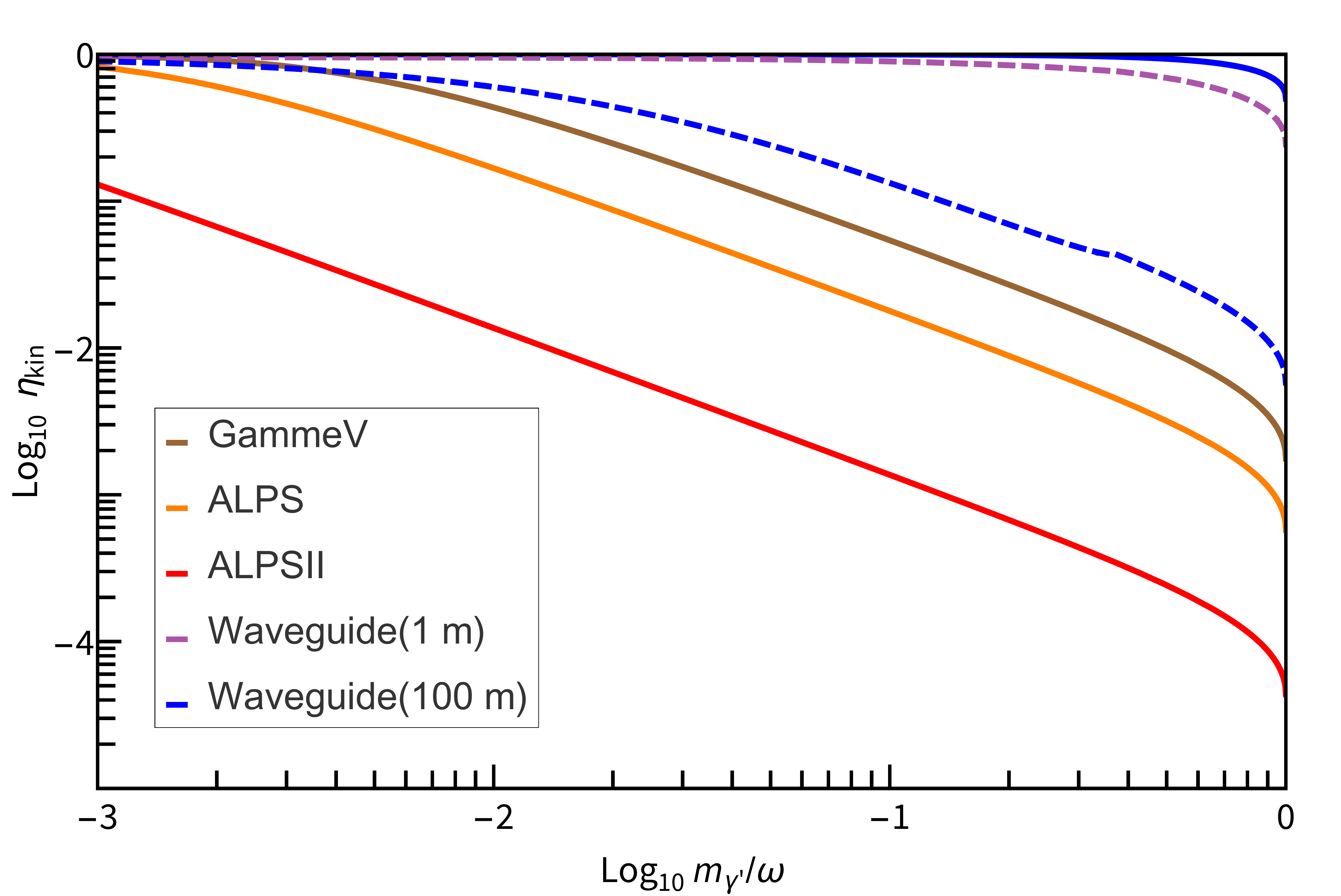}
    \caption{$\eta_\text{kin}$ of the subfrequency new physics signal in various experimental setups. The angular dispersion of the signal photon is large for high dark photon mass, and it weakens $\eta_\text{kin}$ unless a suitable measure is taken (as in waveguide). Dotted curves denote the waveguide with 98.5\% reflectivity coating, while solid ones denote 100\%. The solid blue and solid purple curves (two curves with 100\% reflectivity) coincide. A small kink in the dotted curves are due to the limited wavelength range of the mirror. (See texts in Sec.~\ref{sec:newlaserexperiment} for details.)}
    \label{fig:etakin}
\end{figure}

Let $\ell$ be the distance from the detector to the point a decay occurs.
In our simple setup, we can have
\begin{equation}
\frac{d^2p}{d\ell d\theta_{\text{lab}}} = \frac1L \frac{dp}{d\theta_{\text{lab}}} = \frac1L \frac{dp}{d\theta_{\text{rest}}} \left|\frac{d\theta_{\text{rest}}}{d\theta_{\text{lab}}}\right| .
\end{equation}
The probability that the photon reaches the detector is
\begin{equation}
    \eta_\text{kin}= \int_0^L  d\ell \int_0^{\tan^{-1}(a/\ell)} d\theta_{\text{lab}} \ \frac{d^2p}{d\ell d\theta_{\text{lab}}} ,
    \label{eq:etakin}
\end{equation}
which is straightforward to calculate.

For the later use, we will briefly discuss substitution between $\theta_\text{lab}$ and $E$.  In many cases, calculations become much easier by changing the variable from $\theta_\text{lab}$ to $E$ since $dp/dE$ is uniform. The relations between $E$ and $\theta_\text{lab}$ are given by
\begin{align}
    \theta_\text{lab}(E) & =  \cos^{-1} \frac{2 E \omega -m_{\gamma'}^2}{2E\sqrt{\omega^2- m_{\gamma'}^2}} , \\
    E(\theta_\text{lab}) & = \frac{ m_{\gamma '}^2 \left(\omega + \sqrt{\omega^2-m_{\gamma'}^2 }\cos \theta_{\text{lab}}\right)}{2 \left(\omega ^2 \sin ^2\theta _{\text{lab}}+ m_{\gamma '}^2 \cos^2\theta _{\text{lab}}\right)} .
\end{align}
Using this substitution, Eq.~\eqref{eq:etakin} can be rewritten as
\begin{align}
    \eta_\text{kin} & = \int_0^L  d\ell \int^{E(\theta_\text{lab} = 0)}_{E(\theta_\text{lab} = \tan^{-1}(a/\ell))} \hspace{-5mm} dE \ \frac{d^2p}{d\ell dE} \nonumber \\ & = \frac{1}{\sqrt{\omega^2- m_{\gamma'}^2}} \int_0^L  d\ell \int^{E(\theta_\text{lab} = 0)}_{E(\theta_\text{lab} = \tan^{-1}(a/\ell))}\hspace{-5mm} dE 
    \label{eq:etakin2} \nonumber \\
    & =\hspace{-1mm} \int_0^L\hspace{-1mm}  \frac{m_{\gamma^\prime}^2 \ell (\ell -\hspace{-1mm} \sqrt{\ell^2 + a^2}) + a^2 \omega (\omega + \hspace{-1mm} \sqrt{\omega^2-m_{\gamma^\prime}^2})}{2L(\ell^2 m_{\gamma^\prime}^2 + a^2 \omega^2)}   d\ell .
\end{align}
In Fig.~\ref{fig:etakin}, we show $\eta_\text{kin}$ for various experimental setups.
Some of these setups are the existing LSW experiments that are partly sensitive to our subfrequency new physics scenario.
We used the parameter values specified in Table~\ref{tab:spec}. 
For $m_{\gamma'} \approx \omega$, only a tiny fraction of the signal photon can be detected.
ALPS II has the lowest $\eta_\text{kin}$ due to its smallest $a/L$ ratio.
It is clear how important to collect dispersive signal photons.
As we will detail later in this paper, our new design adopts a waveguide that can fix this, and at least half of the signal photons can reach the detector in ideal case. (See waveguide in Fig.~\ref{fig:etakin}.) 

\begin{table*}[bt]
  \centering
    {%\scriptsize
    \begin{tabular}{c c c c c c c c c}
    \hline\hline
          & a (mm) & $L$ (m) & $\omega$ (eV) & $N_\gamma$ (Hz) & $N_\text{pass}$ & $\eta_\text{eff}$ & $N_d$ (Hz) & $t_s$ (h)\\
    \hline
    Waveguide  & $-$ & $1,\ 100$ & $1.17$ &   $1.6\times 10^{20}$    & $5000$  & $0.54$ & $10^{-6}$ & $480$  \\
    ALPS II & $8.75$\footnote{The ALPS II uses an optic suitable to collect a 17.5 mm diameter beam \cite{Bahre:2013ywa}.} & $100$ & $1.17$ & $1.6 \times 10^{20}$ & $5000$ & $0.95$ & $10^{-6}$ & $480$  \\
    ALPS  & $7$\footnote{Because we are unaware of the radius of the lens, we take the maximum of the vacuum tube size that can be inserted in the HERA dipole magnet \cite{ALPS:2009des}.} & $7.6$ & $2.33$ &      $2.6\times 10^{21}$ & $1$   & $0.9$ & $0.0018$ & $27$  \\
    GammeV & $25.5$\footnote{We take the size of the lens in front of the PMT \cite{GammeVT-969:2007pci}.} & $7.2$ & $2.33$ & $6.6\times 10^{23}$ & $1$   & $0.25$ & $130$ & $24$ \\
    \hline
  SPring-8 & $6$\footnote{We take the radius of crystal in Ge detector \cite{Inada_2013}.} & $0.654$ & $7270 \sim 26000$ & $4.3\times 10^{12} \sim 8.9\times 10^{13} $ & $1$   & $0.23 \sim 0.83 $ & $0.0019 \sim 0.0142$ & $5.28 \sim 8.88$ \\
    \hline \hline
    \end{tabular}
    }
      \caption{Specifications of the experimental setups we use in our analysis. The LSW experiments (GammeV, ALPS, ALPS II) are partly sensitive to the sub-frequency new physics scenario. Our proposed experimental setup (Waveguide) adopts a waveguide. The parameters for the LSW experiments including $\eta_\text{eff}$ were taken from Refs.~\cite{Ehret:2010mh,Flambaum:2019cqi,Bastidon:2015aha,Inada_2013}. Some of the parameter values may not properly reflect the actual experiments. For instance, the $\eta_\text{eff}$ of the LSW experiments would change if the frequency-dependence is properly applied. }
  \label{tab:spec}
\end{table*}

%%%%%%%%%%%%%%%%%%%%%%%%%%%%%%%%%
\section{New laser experiment}
\label{sec:newlaserexperiment}
%%%%%%%%%%%%%%%%%%%%%%%%%%%%%%%%%
\begin{figure}[b]
    \centering
    \includegraphics[width=0.5\textwidth]{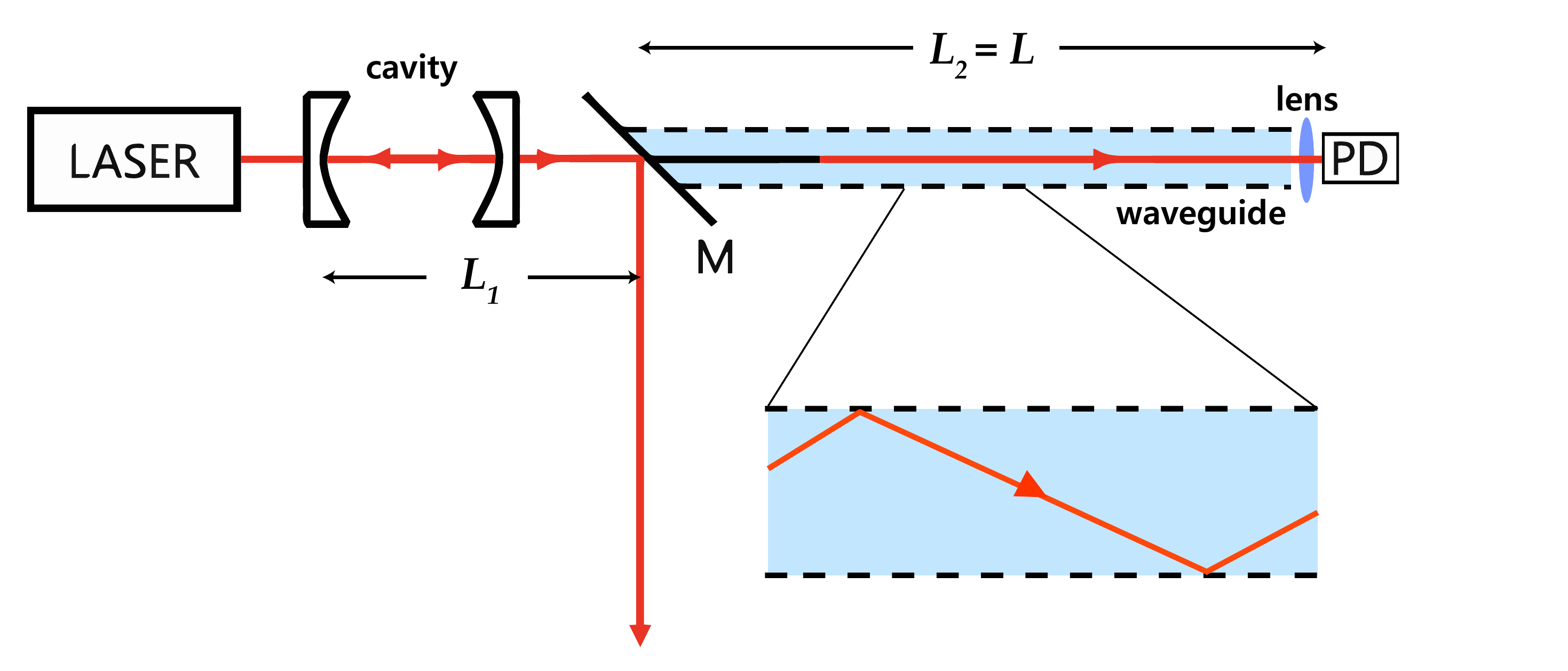}
    \caption{Proposed layout (waveguide). The red is the photon beam, and the black is the dark photon beam. The mirror (M) reflects the laser beam (photon), and only the dark photon can pass the mirror and decay to the subfrequency signal photon. (We do not show the axion in the figure.) The dashed line is the waveguide, which protects the signal from dispersion. A lens is placed in front of the photon detector to focus the signal light into the detector.}
    \label{fig:layout}
\end{figure}

Figure~\ref{fig:layout} shows a schematic diagram of the laser experiment setup that can explore the subfrequency new physics scenario. A small portion of laser light converts to the dark photon, which subsequently decays into the photon and axion.
Because of the different frequencies, we can distinguish the subfrequency signals from the original laser beam in several ways (for instance, we can adopt a prism to distinguish the subfrequency signal light from the original laser light), but here we adopt a rather simple method.
We use a mirror (M) to reflect the laser beam.
Then only the dark photon beam passes the mirror.
The decay can occur anywhere in the propagation of the dark photon, but the decay signal made before the M cannot pass the mirror. Thus, the decay length ($L$) should be taken as the distance between the M and the detector.

The Fabry-Perot cavity before the M is used to amplify the dark photon transition rate through the mirrors, and the cavity amplification ratio ($\eta_\text{cav}$) is given by \cite{Ahlers:2007rd}
\begin{equation}
    \eta_{\text{cav}} = (N_{\text{pass}}+1)/2
    \label{eq:cavinhance}
\end{equation}
where $N_{\text{pass}}$ is the number of the reflections of the laser beam inside the cavity. This depends on the cavity factors and is limited by the damage threshold of cavity mirrors at which the surface coating on the mirror can be damaged. The authors of the Ref.~\cite{Bahre:2013ywa} limit the maximum intensity in the cavity mirror to 500 kW/cm$^2$ (150 kW for 5 mm minimum beam radius in ALPS II). One has a wide choice for the laser intensity and the cavity amplification as long as the circulating power inside the cavity does not exceed the damage threshold. We also take the 150 kW for our maximum circulating power ($\omega N_\gamma N_\text{pass} < 150$ kW).

To avoid losing signal photons from the angular dispersion described earlier, we use a waveguide that can collect and guide all the signal photons in the forward direction toward the detector, which will save at least half of the signals.
A long hollow pipe/rectanguloid with the inside coated by a highly reflective metal would be suitable. The silver-coated mirror has high reflectivity in the optical to near IR range, commercial mirror with more than 98.5$\%$ reflectivity in the 1100 to 20000 nm range is available. (In fact, there is a reduced but still nonzero reflectivity below the 1100 nm wavelength range, but we take it zero for the simplicity.)
We choose the radius of waveguide as the same as that of lens in ALPS II since we also want to use a lens.

In the presence of the waveguide, one should consider the loss of the signal due to the reflections at the mirror. Thus, for the waveguide setup, $\eta_{\text{kin}}$ should be modified to
\begin{equation}
\eta_\text{kin} = \frac{1}{\sqrt{\omega^2-m_{\gamma^\prime}^2}} \int_0^L d\ell \int^{E(\theta_\text{lab}=0)}_{E(\theta_\text{lab}=\pi/2)}  \hspace{-5mm} dE \;  \mathcal{R}(\theta_{\text{lab}} (E),\ell)   \, .
\end{equation} 
The $\mathcal{R}(\theta_{\text{lab}}(E),\ell)$ is a surviving fraction of the photon, and we obtain it as
\begin{equation}
    \mathcal{R}(\theta_{\text{lab}}(E),\ell)= r(E)^{\frac{\ell}{2R} \tan\theta_{\text{lab}}+\frac{1}{2}} \, ,
\end{equation}
where ${R}$ is the radius of the waveguide and $r$ is the reflectivity of the mirror. The power of $r$ is the number of the reflections each signal photon experiences. This expression slightly overcounts the reflection by a decimal, but it is negligible when $r$ is close to $1$.  The $\mathcal{R}(\theta_{\text{lab}},\ell)$ is always less than 1 and $\mathcal{R}(\theta_{\text{lab}},\ell)\rightarrow 1$ as $r \rightarrow 1$.

\begin{figure}[b]
    \centering
    \includegraphics[width=0.45\textwidth]{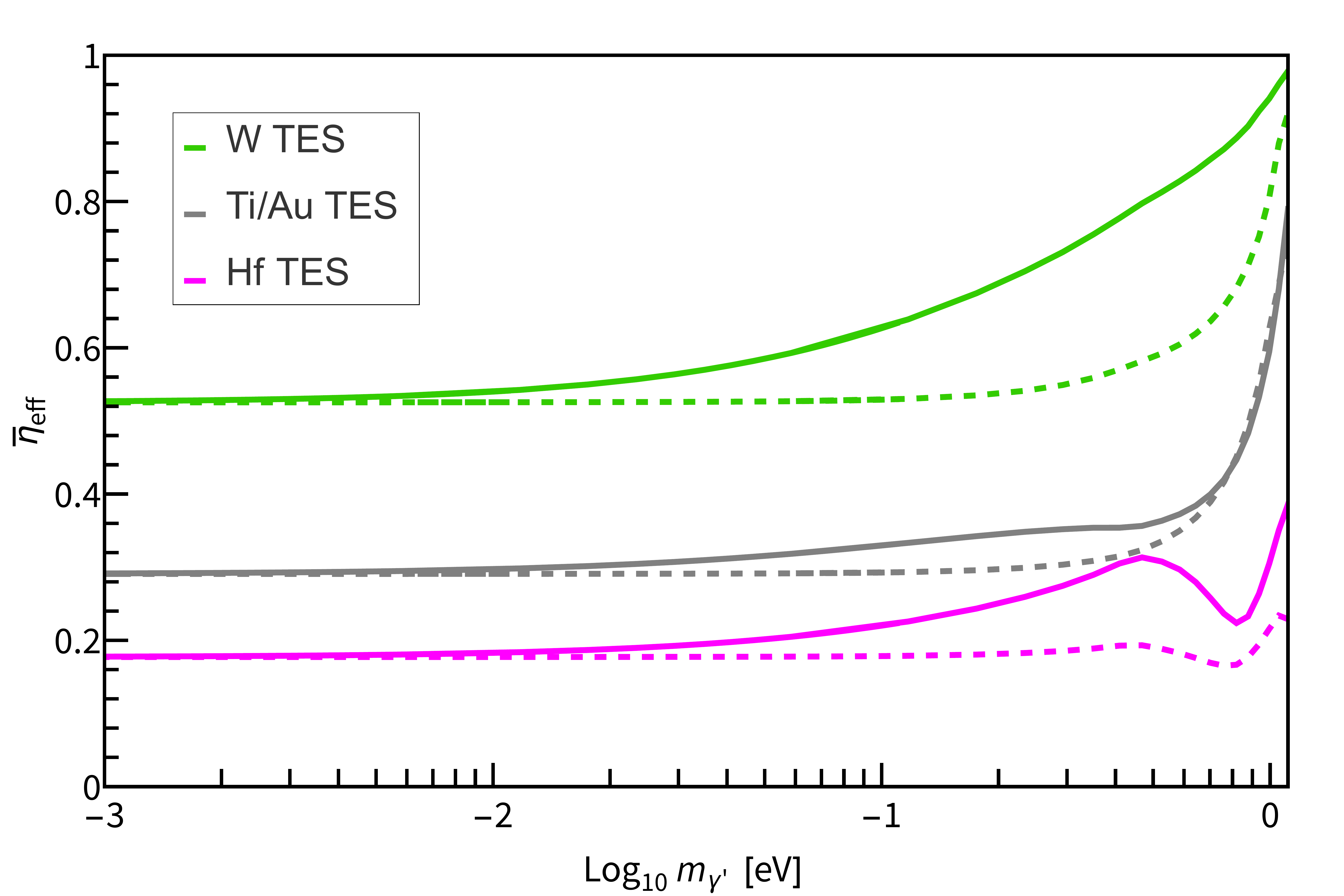}
    \caption{The averaged detector efficiency ($\bar{\eta}_\text{eff}$) with the dark photon mass ($m_{\gamma'}$) for various TES devices when $\omega$=1.17 eV. We take tungsten (W) TES \cite{Lita08}, titanium/silver (Ti/Au) TES \cite{tiudaaa}, hafnium (Hf) TES \cite{ahfawa} for the waveguide with $98.5\%$ reflectivity with $L=1$ m (dashed curves), $L=100$ m (solid curves). Various TES are quite sensitive in the given mass range.}
    \label{fig:etaeff} 
\end{figure}

In general, the SNR of the single photon detection is given by
\begin{equation}
    \text{SNR} = \frac{N_s \sqrt{t_s}}{\sqrt{N_s+N_d}}
    \label{eq:snr}
\end{equation}
where $N_s$ ($N_d$) is the number of signal photon (noise) per second, and $t_s$ is the total measurement time. To be sensitive to small coupling $K$, it is required for the detector to have an extremely low noise. Also, one needs to consider a detector efficiency ($\eta_\text{eff}$), which is a ratio of the detected number of the photon to the incident photon to the detector.

In our subfrequency search scenario, as we discussed in Sec.~\ref{sec:kinematics}, the detector needs to be sensitive to the frequency around $\omega/2$. Thus, the transition edge sensor (TES) which has high sensitivity and low noise in the near IR to the optical range, can be a suitable detector.
Figure~\ref{fig:etaeff} shows the efficiency of TES devices for the waveguide experiment with $L=1$ m and $L=100$ m. Since the detector efficiency mainly depends on the energy of the incident signal, one needs to consider the energy and angular dispersion of the subfrequency signal.

We take the averaged detector efficiency as
\begin{multline}
  \bar{\eta}_\text{eff} = \frac{1}{\eta_{\text{kin}}} \frac{1}{\sqrt{\omega^2-m_{\gamma^\prime}^2}} \\ \int_0^L  d\ell \int^{E(\theta_\text{lab}=0)}_{E(\theta_\text{lab}=\pi/2)} \hspace{-5mm} dE \; \eta_{\text{eff}}(E) \mathcal{R}(\theta_{\text{lab}}(E),\ell)   \, .
\end{multline}
When $m_{\gamma^{\prime}}\ll \omega$, the subfrequency signal is widely dispersed in the energy domain (see Fig.~\ref{fig:pd}).
In that limit, the frequency-wide average makes the sensitivity of each TES flat as shown in Fig.~\ref{fig:etaeff}. In contrast, when $m_{\gamma^{\prime}} \sim \omega$, the subfrequency signals focus at $\omega/2$ and has a large angular dispersion. Therefore, the local shape of the $\eta_{\text{eff}}(E)$ and the signal loss by the waveguide is fed into the curve of $\bar{\eta}_{\text{eff}}$ near the $\omega/2$.

Finally, the ratio of the detectable signal photon number flux to the laser flux is
\begin{equation}
      \frac{N_{s}}{N_{\gamma}}= \eta \frac{N_{\text{sub}}}{N_{\gamma}}=\eta\frac{K^2}{48\pi}\frac{m_{\gamma^{\prime}}^4}{\sqrt{\omega^2-m^2_{\gamma^{\prime}}}}L \, ,
      \label{eq:plt}
\end{equation}
with
\begin{equation}
    \eta\equiv \eta_{\text{kin}}\eta_{\text{cav}}\bar{\eta}_\text{eff} \, .
\end{equation}

%%%%%%%%%%%%%%%%%%%%%%%%%%%%%%%%%
\section{Discussions}
\label{sec:discuss}
%%%%%%%%%%%%%%%%%%%%%%%%%%%%%%%%%
So far, we considered only the case when $m_a \ll m_{\gamma^{\prime}}$. Non-negligible axion mass has the following effects. The decay rate [see Eq.~\eqref{eq:decayrate}] and the energy of the each subfrequency photon decrease as $\Gamma \rightarrow  (1- m_a^2/m_{\gamma^{\prime}}^2)^3\Gamma$, $  E \rightarrow  (1- m_a^2/m_{\gamma^{\prime}}^2)E$. These imply that the subfrequency signal becomes weaker and no longer lies in the near IR to the optical range as $m_a \rightarrow m_{\gamma^{\prime}}$.
This would require microwave or radio frequency detectors, which we do not consider in this paper.

\begin{figure}[tb]
    \centering
    \includegraphics[width=0.45\textwidth]{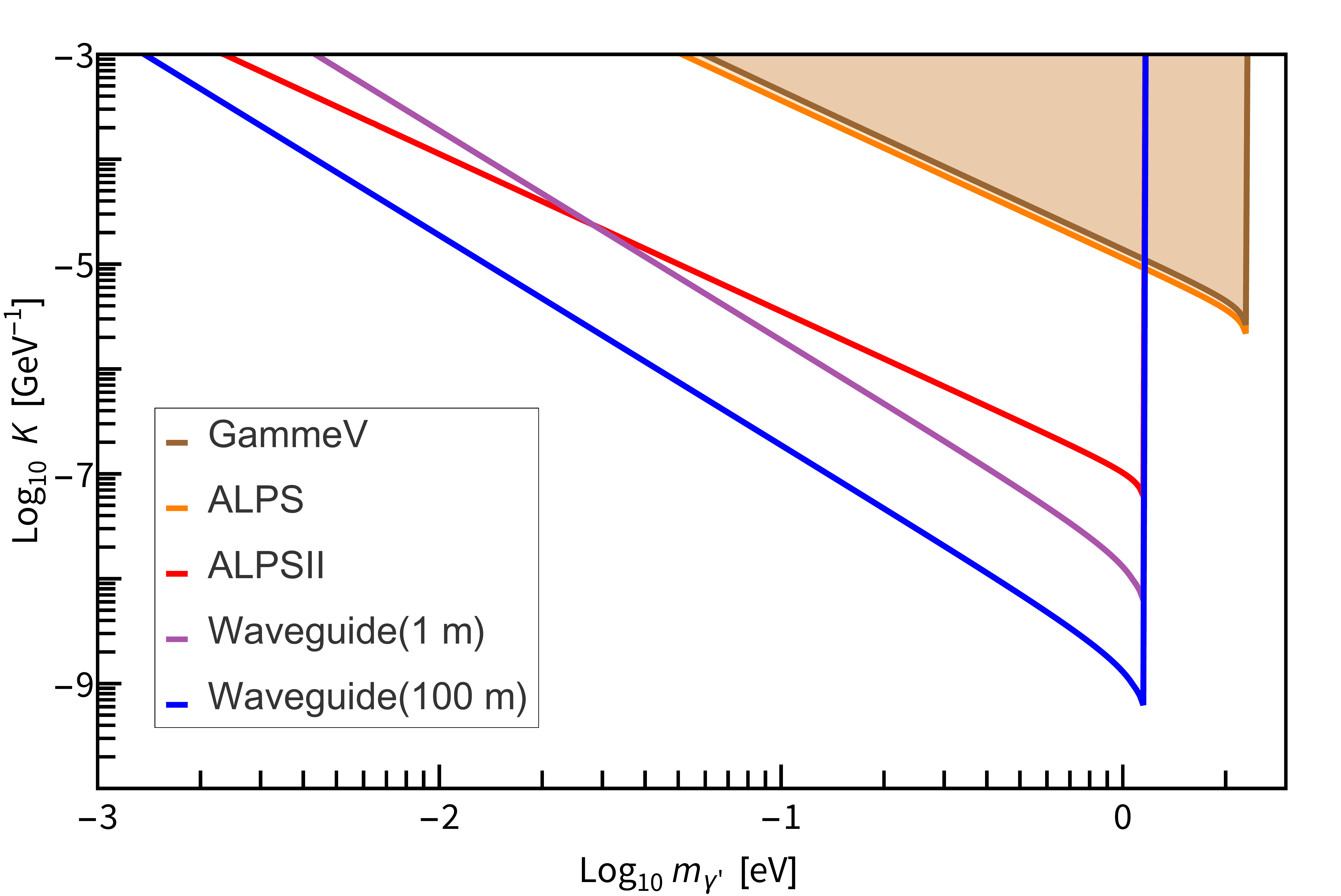}
    \caption{1$\sigma$ C.L. exclusion/sensitivity limits in the mass ($m_{\gamma'}$) and coupling ($K$) parameter space for various experimental setups.
    For waveguides, we chose the same optical parameters and the measurement time ($t_s=20$ days) as the ALPS II, which is the state-of-the-art experiment for the high-power, low-signal measurement (see the parameters and caveats in Table~\ref{tab:spec}). }
    \label{fig:sensitivity} 
\end{figure}

\begin{figure}[b]
    \centering
    \includegraphics[width=0.45\textwidth]{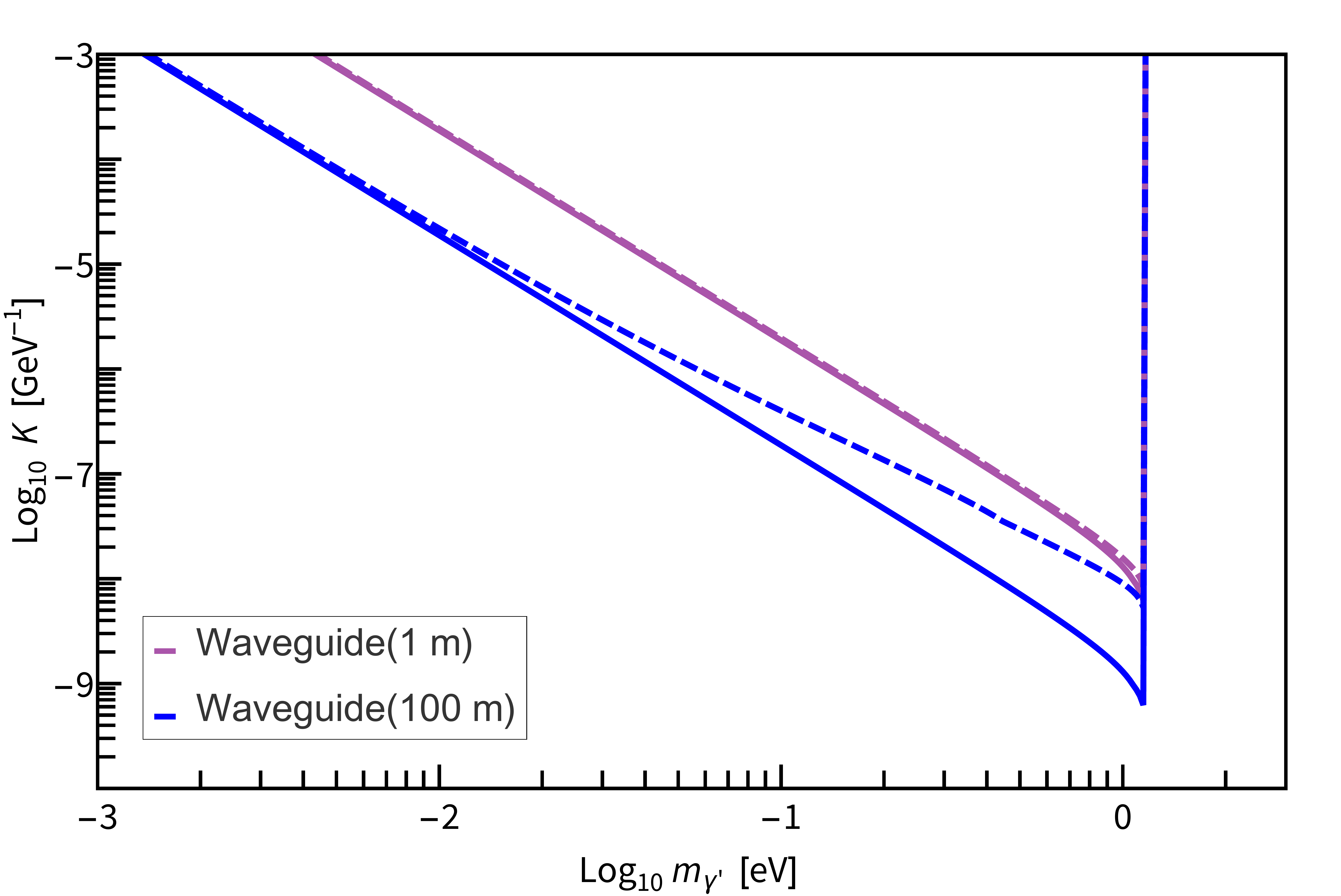}
    \caption{A comparison of 1$\sigma$ C.L. sensitivity limits for the waveguides of the $98.5\%$ (dashed curves) and $100\%$ (solid curves) reflectivity coating. The signal loss from the reflection is more critical when the length of waveguide is larger, since the number of the reflection is proportional to the length.} 
    \label{fig:wgloss} 
\end{figure} 

Figure~\ref{fig:sensitivity} shows the exclusion limits and the future sensitivities of various experimental setups. The waveguide sensitivity curves are for the ideal, 100$\%$ reflectivity case. The comparison to the realistic case of the $98.5\%$ reflectivity is shown in Fig.~\ref{fig:wgloss}.
We made a cutoff for each curve at $m_{\gamma'} = 0.99 \omega$. This is far below the spectrum of the laser due to its linewidth.
Also, the limit \eqref{eq:Limit} is satisfied by this cutoff. 
Since the decay rate of the dark photon (thus the signal) increases with its mass, the experiment is particularly sensitive when the $m_{X}$ is close to the laser frequency $\omega$, which can be seen from the negative slopes of curves. In contrast, the subfrequency signal is more dispersive in the angular domain as $m_{\gamma^{\prime}}\rightarrow\omega$. This reduces the signal reaching the detector (see Fig.~\ref{fig:etakin}).

The new setup with a waveguide can excel the early LSW experiments (GammeV, ALPS) by orders of magnitude.
Even tabletop experiments ($L = 1$ m) can outperform them.
ALPS II is an ongoing LSW experiment, which has the most extended baseline ($L = 100$ m).
This long baseline clearly improves its sensitivity, although it is not using a waveguide. However, the ALPS II gets the most stringent degradation from the angular dispersion due to its large baseline, as one can see from its gentle slope.
As the curves show, using a waveguide for a given $L$ significantly improves the sensitivity, especially in the high mass region.

A rough estimate for the QCD axion can be done, if one assumes $G_{a\gamma\gamma} \sim G_{a\gamma\gamma^{\prime}}$. The experimental/observational bound is $G_{a\gamma\gamma} < 10^{-10} $ GeV$^{-1}$ for the QCD axion of lighter than the eV scale \cite{ParticleDataGroup:2020ssz}. Our sensitivity study in $K$ shown in Fig.~\ref{fig:sensitivity} indicates the QCD axion cannot be detected with the current technology even if we take the largest possible kinetic mixing $\varepsilon \sim 1$.

The sensitivity would further increase if we can (i) increase the circulating laser power, which is currently limited by the damage threshold of the mirrors in the cavity (i.e., improve the mirror coating), (ii) reduce the detector background noise, (iii) extend the baseline for the dark photon decay (accompanied by a waveguide).
It can also reach a higher mass using the higher frequency light source.
(For instance, see the mass coverage of GammeV and ALPS, which use relatively high-frequency lasers.)

\begin{figure}[tb]
    \centering
    \includegraphics[width=0.45\textwidth]{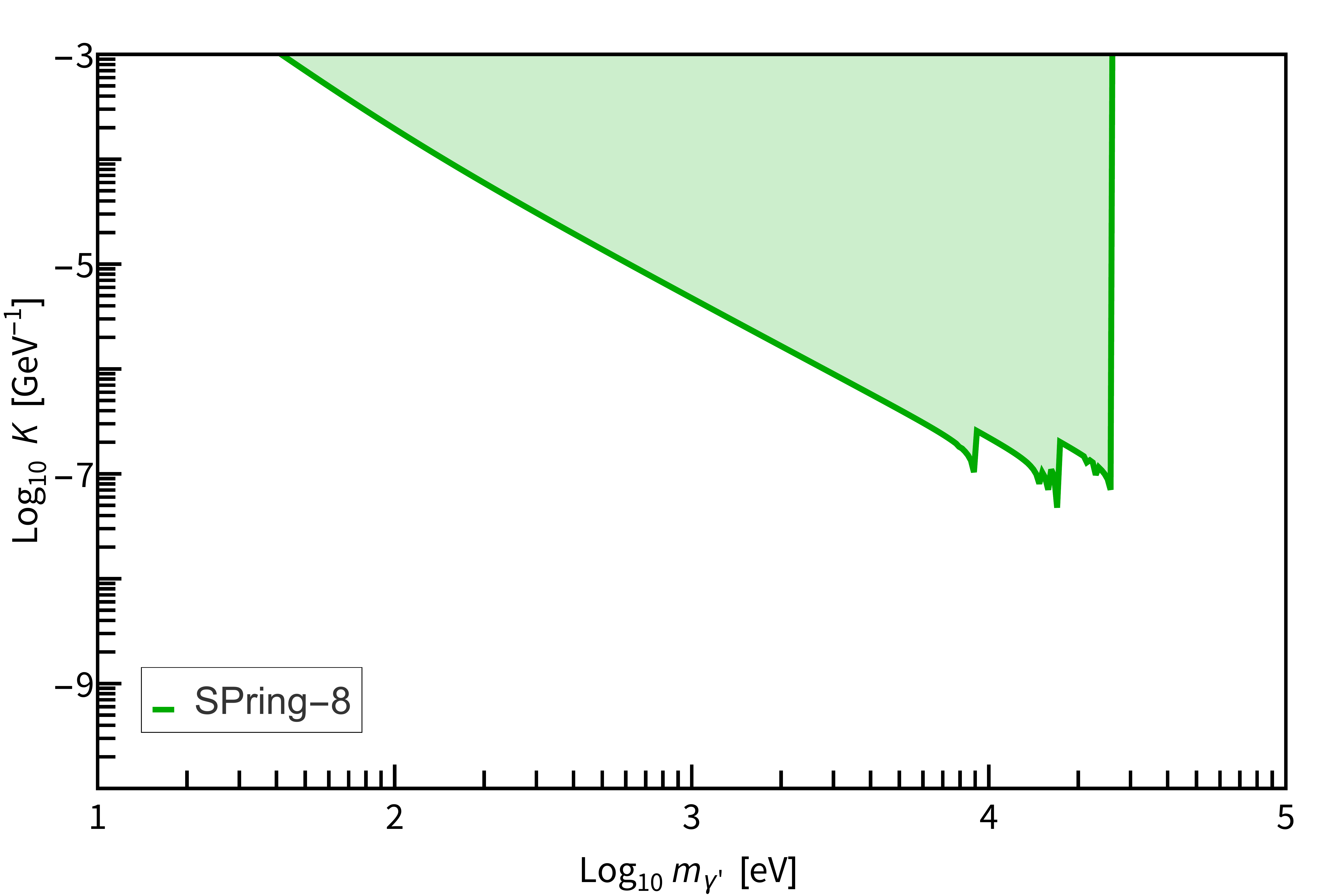}
    \caption{1$\sigma$ C.L. exclusion limits in the mass ($m_{\gamma'}$) and coupling ($K$) parameter space for SPring-8 experiments \cite{Inada_2013}. SPring-8 experiments consist of a series of LSW experiments with various frequency x-ray sources. The sawtooth shape comes from the combination of a few experiments.} 
    \label{fig:spring} 
\end{figure}

Especially, an x-ray source can cover high mass region where the decay rate is large.
In Fig.~\ref{fig:spring}, we show the constraints from the null result of the light shining through wall experiment SPring-8 using the x-ray \cite{Inada_2013}.
Although a waveguide cannot be not used for the x-ray due to the high transmittance, it already gives a sizable constraints.

One of the critical differences of the proposed experiment from the other laser-based new physics searches is that it exploits the decay of a new particle.
Because of this property, it results in several differences such as the angular deviation of the signal, need of sizeable decay length, no need of wall in principle (we could use a prism instead of the wall), and different signal frequency from the original laser frequency.
These distinctions allow a different way of optimization from the LSW, such as the waveguide we take.

%%%%%%%%%%%%%%%%%%%%%%%%%%%%%%%%%
\section{Summary and outlook}
\label{sec:summary}
%%%%%%%%%%%%%%%%%%%%%%%%%%%%%%%%%
Our study shows a simple setup of an eV-scale optical laser and a photon detector can do the entire process from the production of a new particle, its decay into the SM particles, and the new physics signal detection.
It was considered possible only in high-energy accelerator experiments not long ago.
It is striking to note the whole setup can be as small as a tabletop experiment of a 1-meter scale.

One of the crucial points is that the test of the combined portal $K = \varepsilon G_{a\gamma\gamma'}$ cannot be obtained just by multiplying the bounds on $\varepsilon$ and $G_{a\gamma\gamma'}$ obtained independently.
For instance, if we use only $\varepsilon$, the dark photon lighter than 1 MeV would not practically decay \cite{Pospelov:2008jk}.
Also, the $\varepsilon$ bound from the stellar objects can be weakened due to the reconversion of the dark photon with the combined portal.
If we use only $G_{a\gamma\gamma'}$, on the other hand, then the dark photon production will be greatly suppressed because it relies on an off-shell photon.
Therefore, we need a dedicated study for the implications and constraints of the combined portal $K$.
In this sense, it can be taken as a practically new portal, which we name the ``subfrequency portal.''

The essential part of this portal is that any source of light can produce subfrequency light as a signal of new physics.
While we focused on the new physics search in the laser experiments, which cover the visible light frequency, plus some x-ray examples, there are plenty of possible extensions such as microwave cavities and isotope spectrum spectroscopy. Also, the astrophysical/cosmological data from the stellar objects and cosmic microwave backgrounds are wonderful places to study.
Rich physics implications are warranted.

\begin{acknowledgments}
J. L. thanks J. Jung and H. Kim for helpful conversations about the optical devices. 
This work was supported in part by the National Research Foundation of Korea (Grant No. NRF-2021R1A2C2009718).
\end{acknowledgments}

\bibliography{ref.bib}

\end{document}